\documentclass[aps,prb,groupedaddress,twocolumn,superscriptaddress]{revtex4-1}
\usepackage{enumitem}

\usepackage{graphicx} % Include figure files
\usepackage{bpchem}
\usepackage{dcolumn}% Align table columns on decimal point
\usepackage{bm}% bold math
\usepackage{amsmath,amsthm,amssymb}
\usepackage{color}
\usepackage{hyperref}% add hypertext capabilities
\usepackage{multirow}
\usepackage{array}
\newcommand{\cuse}{Cu$_2$OSeO$_3$}
\usepackage{hhline}
\makeatletter
\renewcommand*{\@fnsymbol}[1]{\ensuremath{\dagger}}
\makeatother

\begin{document}
\title{Dissipation phenomena and magnetic phase diagram of \cuse~below 50~K}
\author{F. Qian}
\affiliation{Faculty of Applied Sciences, Delft University of Technology, Mekelweg 15, 2629 JB Delft, The Netherlands}
\author{H. Wilhelm}
\affiliation{Diamond Light Source Ltd., Chilton, Didcot, Oxfordshire, OX11 0DE, United Kingdom}
\author{A. Aqeel}
\affiliation{Zernike Institute for Advanced Materials, University of Groningen, Nijenborgh 4, 9747 AG Groningen, The Netherlands}
\author{T.T.M. Palstra}
\thanks{Present Address: RM University of Twente, The Netherlands}
\affiliation{Zernike Institute for Advanced Materials, University of Groningen, Nijenborgh 4, 9747 AG Groningen, The Netherlands}
\author{A.J.E. Lefering}
\author{E.H. Br\"uck}
\author{C. Pappas}
\affiliation{Faculty of Applied Sciences, Delft University of Technology, Mekelweg 15, 2629 JB Delft, The Netherlands}

\date{\today ~~susc-lowT$_-$cuse.pdf}

\begin{abstract} 
\noindent  
The magnetic phase diagram of the chiral magnet \cuse~has been investigated by dc magnetisation and ac susceptibility over a very large frequency range from 0.1~Hz up to 1~kHz and temperatures below 50~K. Qualitatively different phase diagrams have been obtained by applying the magnetic field along the easy $\langle100\rangle$ or the hard $\langle110\rangle$ crystallographic directions. Strong $\chi''$ appears close to $B_{c2}$, but only below 30~K and is more pronounced when $B$ is applied along $\langle100\rangle$. In addition, the transition from the helical to the conical phase leads to two adjacent $B_{c1}$ lines below 50~K. The frequency dependence of $\chi''$ indicates the existence of very broad distributions of relaxation times  both at $B_{c1}$ and $B_{c2}$. The associated very long characteristic times eventually prevent the system from reaching thermal equilibrium at low temperatures and lead to thermal hysteresis. 

%The borders between the helical and the conical phases at $B_{c1}$, and between the conical and the field polarised phases at $B_{c2}$ are identified both from the real ($\chi'$) and imaginary ($\chi''$) components of the ac susceptibility. Magnetic dissipation phenomena probed by $\chi''$ reflect the existence of very broad distributions of relaxation times, which eventually prevent the system from reaching thermal equilibrium at low temperatures and lead to thermal hysteresis at both $B_{c1}$ and $B_{c2}$. When $B$ is applied along the easy axis $\langle100\rangle$, both $\chi'$ and $\chi''$ show marked peaks close to $B_{c2}$, which by comparison with theory give a hint for a smeared first order transition from the conical to the field polarised phases for this sample orientation with respect to the magnetic field.
\end{abstract}

\maketitle

\section{Introduction}
Magnetic chirality generated by the anti-symmetric Dzyaloshinskii-Moriya (DM) interaction is at the focus of interest since the observation of stable skyrmion lattice textures in systems of the B20 family, such as MnSi \cite{Muhlbauer:2009bc, Tonomura:2012ep}, Fe$_{1-x}$Co$_x$Si \cite{Yu:2010hr, 2010PhRvB..81d1203M} or FeGe \cite{Yu:2010iu, Moskvin:2013kf}. The existence of these particle-like objects in chiral magnets was predicted in the late 1980's by Bogdanov and Yablonskii \cite{Bogdanov:1989tt, Bogdanov:1989uw} and their topological robustness is at the origin of several new phenomena paving the way to potential applications in data storage and manipulation devices \cite{Fert:2013fq}. 

%Skyrmions are fundamentally interesting because their topological stability particle-like solutions structure gives rise to effective electromagnetic fields acting on electrons and magnons Skyrmions, introduced in field theory b  Skyrme \cite{Skyrme:1961vo} in the early 1960's are becoming an important   in the late 1980's by Bogdanov and Yablonskii  \cite{Bogdanov:1989tt, Bogdanov:1989uw} to chiral magnetism, where  they may even occur spontaneously  \cite{Rossler:2006cq}. are becoming  an important concept. These topologically protected objects are stabilised in chiral magnets and 

The multiferroic insulator \cuse~is  a chiral magnet hosting skyrmions,  that can be manipulated by electric fields \cite{White:2014jia, Seki:2012ie}. This system has a similar phase diagram as the metallic B20 compounds, determined experimentally by several studies \cite{Adams:2012du,Zivkovic:2012kr, Belesi:2012fi,Seki:2012ie,Seki:2012ch,Omrani:2014iea,Levatic14,Qian2016} close to $T_c=58.2$~K. At zero field, the helix has a pitch of $\sim$60~nm arising from the balance between the strongest ferromagnetic exchange  and the DM  interactions and is fixed by a weak magnetocrystalline anisotropy along the $\langle100\rangle$ crystallographic directions. Relatively weak external magnetic fields exceeding $B_{c1}$ orient the helices along their direction, leading to the conical phase. High enough fields above $B_{c2}$ overcome the DM interactions and induce a field-polarised  state. The skyrmion lattice is stabilised at intermediate fields between $B_{c1}$ and $B_{c2}$ in a narrow temperature window just below $T_c$. 

\cuse~is characterised by a more complex crystal structure than the metallic B20 compounds, although with the same P2$_1$3 space group. The unit cell contains 16 Cu$^{2+}$ ions on two different sites, Cu(1) and Cu(2) \cite{Bos08, 2014PhRvB.89n0409D}, which form tetrahedra composed of three Cu(1) and one Cu(2). In contrast to the B20 metallic chiral magnets, in \cuse~the superexchange magnetic interactions between the Cu$^{2+}$ spins and the DM interactions can be calculated ab-initio  at the atomic level \cite{2012PhRvL.109j7203Y, Janson:2014uo, 2014PhRvB..90n0404R} providing a direct link between theory and experiment. This approach reveals the existence of weak and strong magnetic bonds and a separation of exchange energy scales beyond the one encountered in the B20 metallic counterparts \cite{Bak:1980ud}. 

In \cuse~the  building blocks of the magnetic structure are the  Cu$^{2+}$ tetrahedra, which  interact weakly with each other \cite{Janson:2014uo, 2014PhRvB..90n0404R}. Within the tetrahedra the spin of the single  Cu(2) ion is  antiparallel to the spins of the three Cu(1) ions, resulting in a total spin of S=1 and a magnetic moment of  $\sim$0.5 $\mu_B$ per Cu$^{2+}$ ion. This description is consistent with the quantum mechanical treatment of the S=1/2 spins  and has profound implications on the magnetic properties. In particular, the low temperature phase diagram of \cuse~ is qualitatively different when a magnetic field  is applied along the easy  $\langle100\rangle$ crystallographic direction or not \cite{Janson:2014uo}. 
 
%which shows a glassy behavior as seen in electron microscopy \cite{Rajeswari}. ........

In this study we investigate the $B-T$ phase diagram of \cuse~ at temperatures below 50~K by dc magnetisation and ac susceptibility measurements spanning a large frequency range from  0.1 Hz up to 1~kHz. For a  comparison with theory the magnetic field has been applied along the easy  $\langle100\rangle$  and the hard  $\langle110\rangle$  crystallographic directions. The $B_{c1}$ transition from the helical to the conical phase splits in two  distinct lines below 50~K, which may be related to the two different Cu$^{2+}$ sites. The transition to the field polarised state at $B_{c2}$ is seen not only on the real component of the susceptibility, $\chi'$, but also on the imaginary one, $\chi''$, most prominently for $B \parallel \langle100\rangle$.  The weak frequency dependence of $\chi''$ and  thermal hysteresis phenomena at both $B_{c1}$ and $B_{c2}$ reflect the existence of  long relaxation times with broad distributions. The results are compared with theory, which  predicts more dramatic changes with a first order phase transition at $B_{c2}$ for $B\parallel\langle100\rangle$ and does not account for the two $B_{c1}$ lines observed. 

%############################################################################
%    figures
%   MvsB.pdf#
%############################
\begin{figure*}
\includegraphics[width= 0.79\textwidth]{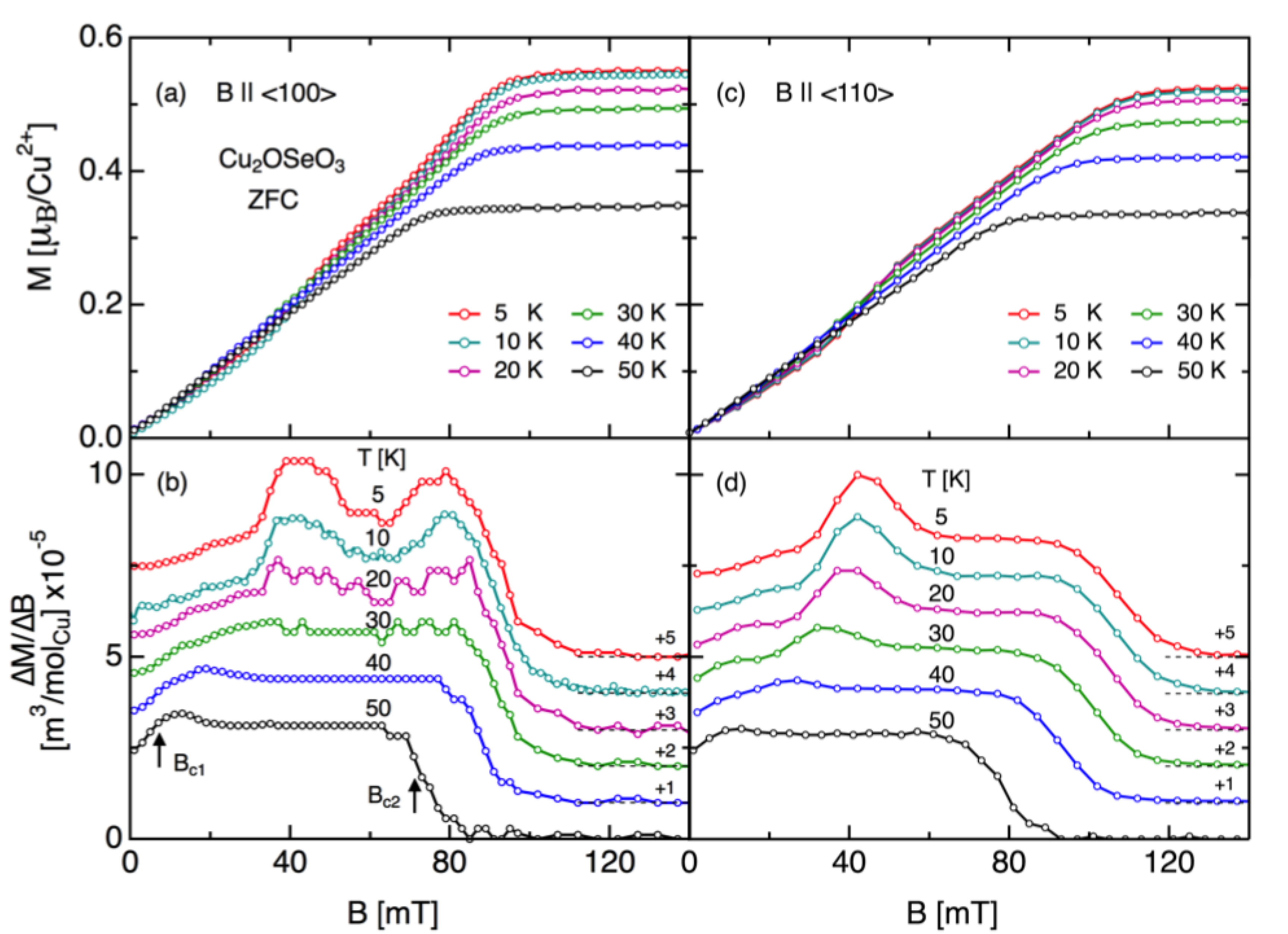}
\caption{Magnetic field dependence of the ZFC magnetisation, panels (a) and (c), and of the deduced susceptibility $\Delta M/\Delta B$, panels (b) and (d), of \cuse~between 5 and 50~K for $B\parallel\langle100\rangle$, panels (a) and (b), and $B\parallel\langle110\rangle$, panels (c) and (d). The arrows to the 50~K curve in  (b) indicate the critical fields $B_{c1}$ and $B_{c2}$ determined from the  inflection points of $\Delta M/\Delta B$. For the sake of clarity the $\Delta M/\Delta B$ curves have been shifted vertically with respect to the base line as indicated.} 
\label{Fig:MvsB.pdf}
\end{figure*}
%#########################################################################
%############################################################################
%    figures
%   MvsT.pdf#
%############################
\begin{figure*}
\includegraphics[width= 0.79\textwidth]{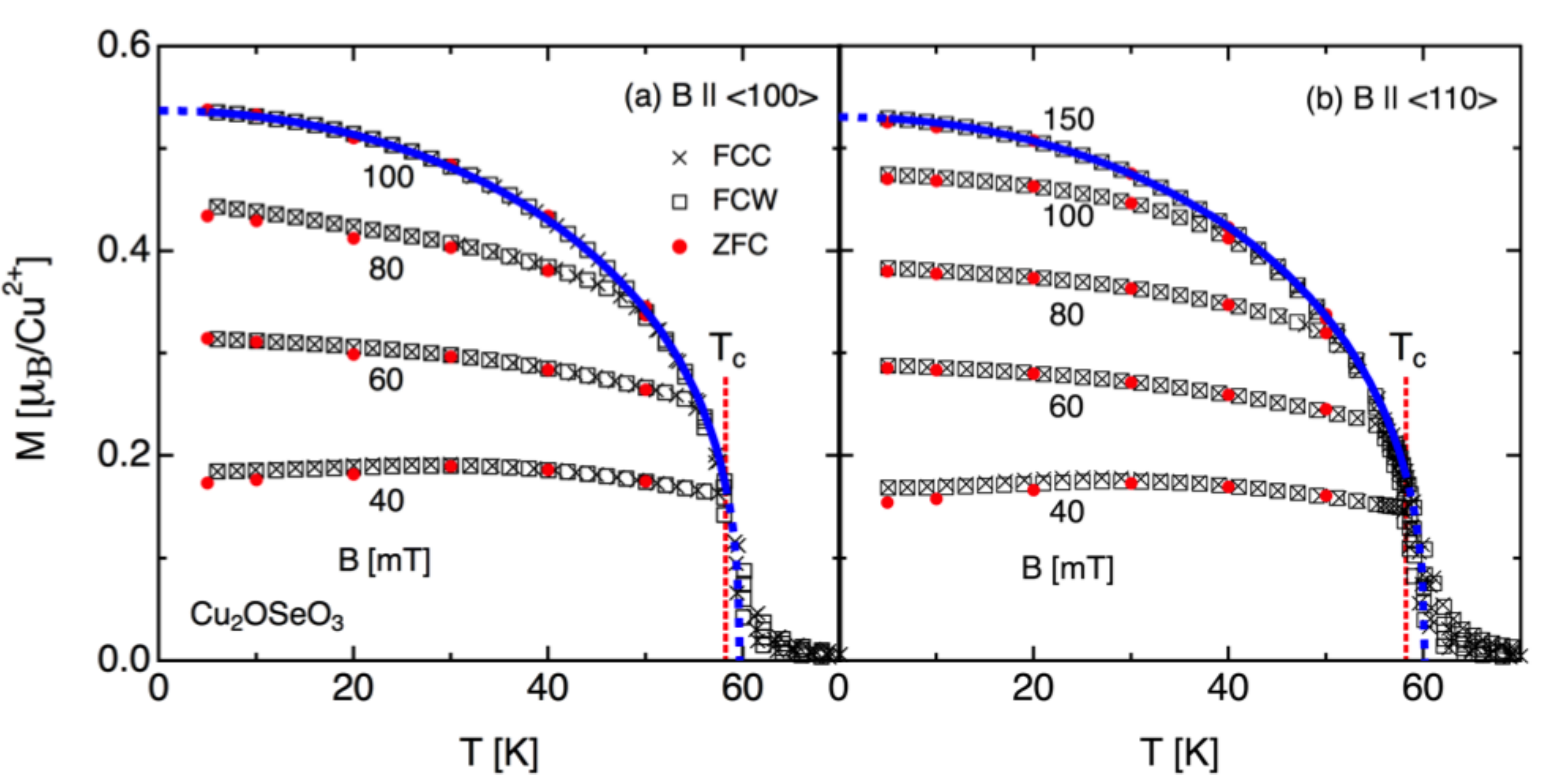}
\caption{Temperature dependence of FCC and FCW magnetisation of \cuse~at several magnetic fields for (a) $B\parallel\langle100\rangle$ and (b) $B\parallel\langle110\rangle$. The ZFC data from $M$ vs $B$ measurements are included as well. The blue lines in both panels are fits of eq.~\ref{eq:msat} with the parameters given in the text. The red vertical dashed lines indicate $T_c=58.2$~K.} 
\label{Fig:MvsT.pdf}
\end{figure*}
%#########################################################################
%############################################################################
%    figures
%   chivsB.pdf#
%############################
\begin{figure*}
\includegraphics[width= 0.78\textwidth]{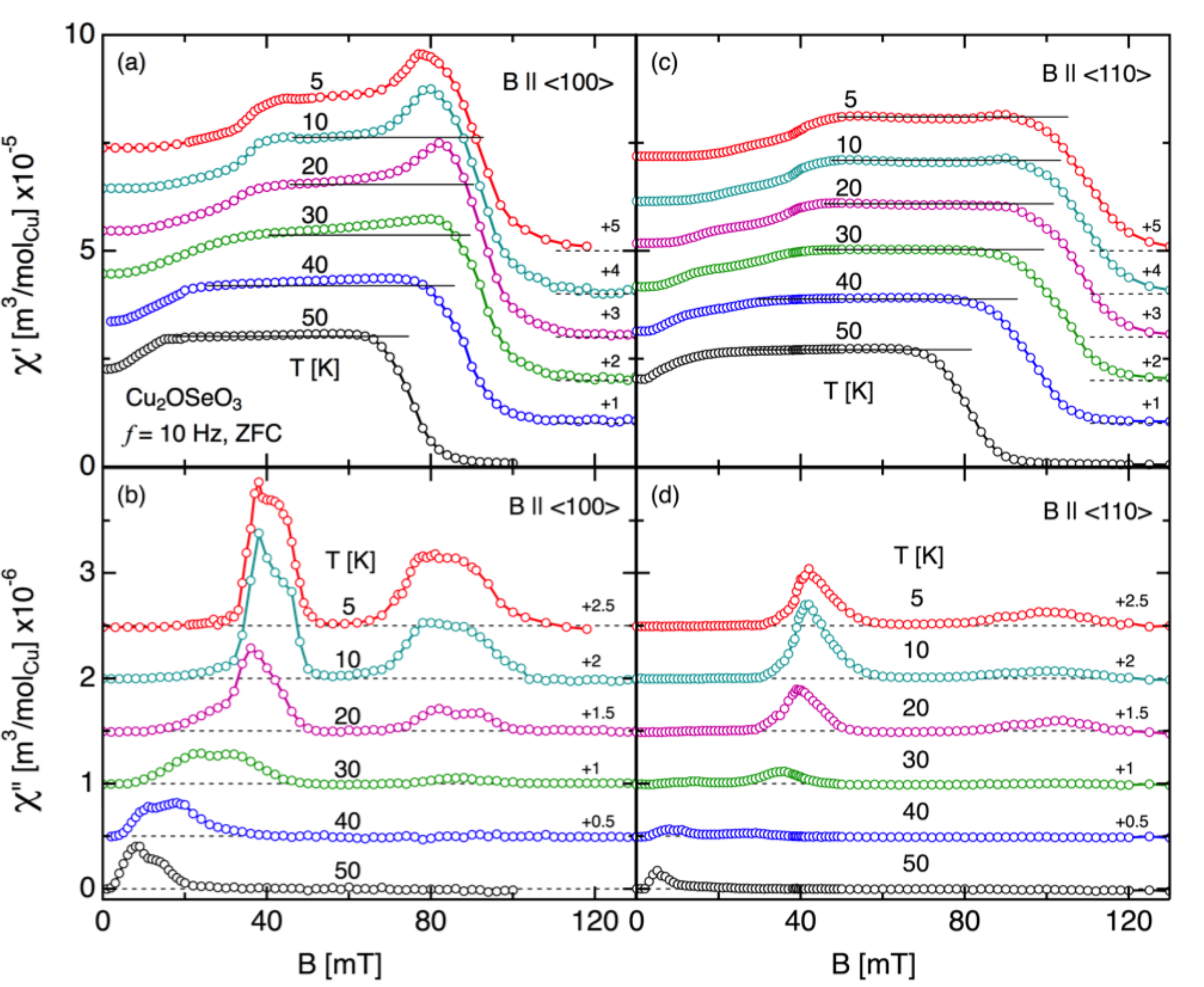}
\caption{Magnetic field dependence of ZFC $\chi'$ and $\chi''$ of \cuse~at selected temperatures from 5 to 50~K at $f=10$~Hz.  Panels (a) and (b) correspond to $B\parallel\langle100\rangle$ and panels (c) and (d) to $B\parallel\langle110\rangle$. The horizontal  lines in panels  (a) and (c) are guides to the eye. All curves have been shifted vertically with respect to the base line as indicated.} 
\label{fig:chivsB.pdf}
\end{figure*}
%############################################################################
%############################################################################
%    figures
%   chivsT.pdf#
%############################
\begin{figure}
\includegraphics[width= 0.4\textwidth]{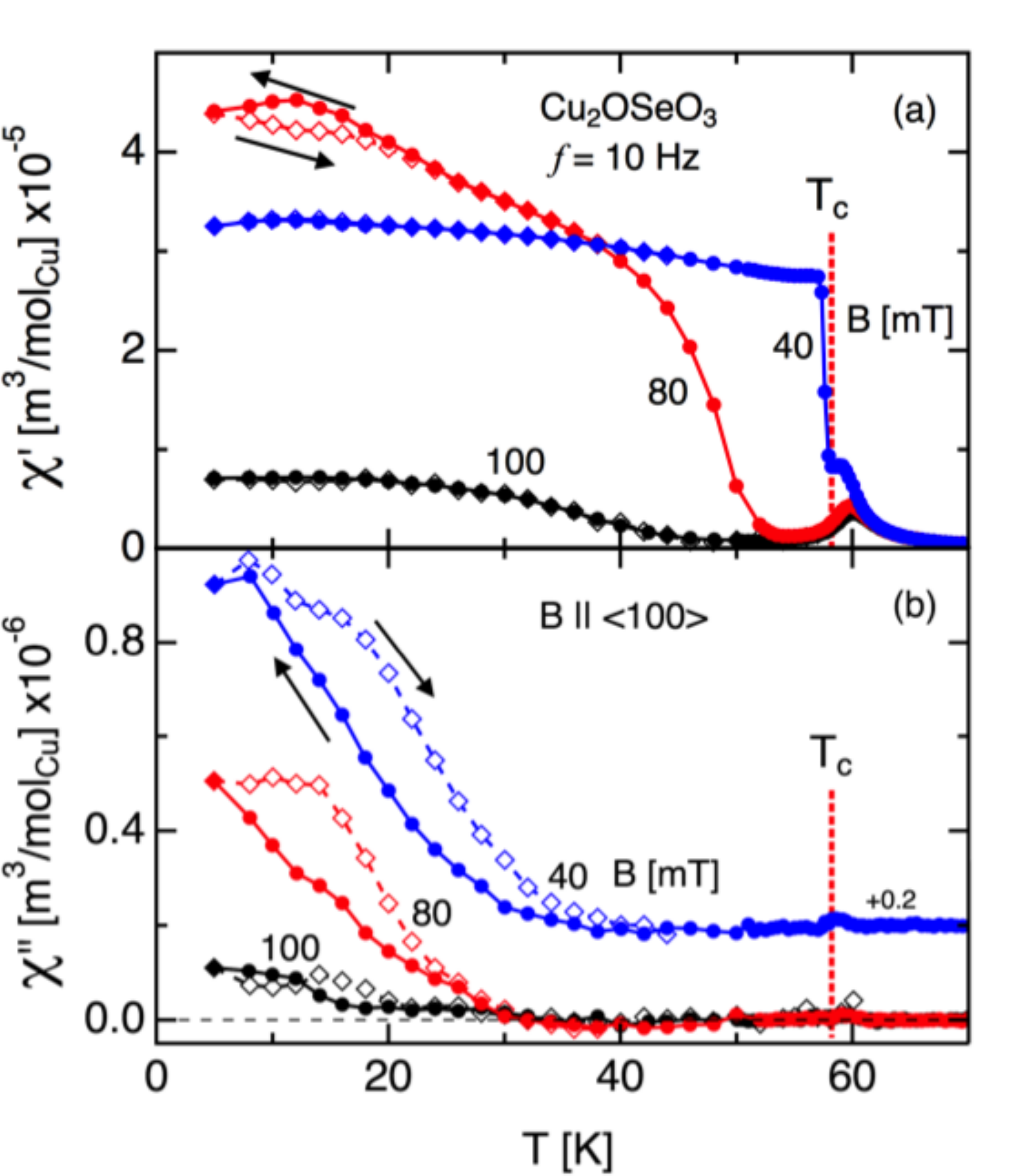}
\caption{Temperature dependence of FCC (solid markers) and FCW (open markers) $\chi'$ (a) and $\chi''$ (b) of \cuse~at various magnetic fields and $f=10$~Hz for $B\parallel\langle100\rangle$. The vertical lines indicate $T_c=58.2$~K. The $\chi''$ curve for 40~mT has been shifted vertically with respect to the base line as indicated. } 
\label{fig:chivsT.pdf}
\end{figure}
%############################################################################

%############################################################################
%############################################################################
\section{Experiment details}
Two single crystals of \cuse~with dimensions of $\sim1\times1\times1$ mm$^3$ were prepared by chemical vapour transport method \cite{Aqeel2016,Miller:2010fu} and their structure  was checked by single crystal x-ray diffraction. One sample was oriented with $\langle100\rangle$ vertical and was the same as in the previous study \cite{Qian2016}. The other sample was oriented with $\langle110\rangle$ vertical. The accuracy of the orientation for both samples was within $\pm5^\circ$. 

The measurements were performed with a MPMS-5XL superconducting quantum interference device (SQUID) using the extraction method. The dc magnetic field $B$ was applied along $\langle100\rangle$ or $\langle110\rangle$, as was the drive ac field $B_{ac}=$ 0.4~mT with frequency ranging from 0.1 to 1000~Hz. The measurements yield the dc magnetisation $M$ as well as the real and imaginary components of the ac susceptibility, $\chi'$ and $\chi''$ respectively. Three specific experimental protocols were adopted in order to investigate the influence of magnetic history:
%\begin{compactitem} [noitemsep]
\begin{itemize}[noitemsep, leftmargin=*, topsep=0pt]
\item  Zero-field-cooled scans (ZFC): the sample was brought to the temperature of interest in zero field and then the measurements were performed by increasing stepwise the magnetic field. 
\item  Field-cooled cooling scans (FCC): the sample was brought to 70~K, a magnetic field was applied and the signal was recorded by decreasing stepwise the temperature.
\item  Field-cooled warming scans (FCW): the sample was brought to 70~K, and then cooled down to 5 K in a magnetic field. The signal was recorded by increasing the temperature stepwise.
\end{itemize}
All measurements have been performed once the sample has reached thermal equilibrium. 
%=====================================================
% 			Results outline
%			chi' and chi'' vs B at 10 Hz for 50, 40, 30, 20, 10, 5 K
%			chi' and chi'' vs T at 10 Hz for low and high B
%			chi' and chi'' vs B for various f at 40 K and 5 k
%			thermal hysteresis chi' and chi'' vs T at low and high fields for various f
%====================================================
%====================================================
\section{dc magnetisation}
\label{sec:dc}
Figure~\ref{Fig:MvsB.pdf} displays the magnetic field dependence of the ZFC magnetisation $M$ and its corresponding susceptibility $\Delta M/\Delta B$, deduced from numerical differentiation, for a set of temperatures between 5 and 50~K. The results are given for $B$ along  $ \langle100\rangle$, thus along the easy axis of the magnetic structure, in panels (a) and (b), and for $B$ along $\langle110\rangle$, in panels (c) and (d). 

For both orientations, $M$ increases almost linearly with $B$ at low fields, which leads to a slowly varying $\Delta M/\Delta B$. By further increasing the magnetic field, a kink appears in the magnetisation curves, which is more clearly seen as a peak in the derivative $\Delta M/\Delta B$  and marks the transition from the helical to the conical phase at the critical field $B_{c1}$. For both sample orientations this peak is more pronounced at low temperatures and at 5~K it shifts to $\sim$40~mT  in agreement with the $B_{c1}$ values reported in the literature \cite{Bos08,Huang:2011uf,Adams:2012du,Belesi:2012fi}. 

At high fields, $M$ levels off above $B_{c2}$ and reaches a temperature dependent value $M_{sat}$, which at 5~K amounts to $\sim$0.54~$\mu_B$/Cu$^{2+}$ for both sample orientations in agreement with literature \cite{Bos08,Huang:2011uf, Zivkovic:2012kr, Belesi:2012fi, Adams:2012du}. However, the way $M$ approaches $M_{sat}$ below 20~K is very different for the two sample orientations. This is clearly seen in the deduced $\Delta M/\Delta B$ curves, where only for $B \parallel \langle100\rangle$ a maximum, centered at $\sim$80~mT for 5~K, develops close to $B_{c2}$. The observed behaviour and the differences between the two sample orientations follow the same trend as reported in the literature \cite{Belesi:2012fi}.

% and recently theoretically predicted magnetisation curves \cite{Janson:2014uo} which however shows much more dramatic differences with a continuous smooth transition for $B \parallel \langle111\rangle$ and a discontinuous one involving a jump in the magnetisation and a  divergence of the derivative at $T=0$~K, at  $B_{c2}$ for $B \parallel \langle100\rangle$.

Figure~\ref{Fig:MvsT.pdf} displays the temperature dependence of the FCC and FCW magnetisations for some specific magnetic fields with $B\parallel \langle100\rangle$ in panel (a) and $B\parallel \langle110\rangle$ in panel (b). The ZFC data from the $M$ vs $B$ curves are included as well. For both orientations, the ZFC magnetisation is slightly lower than the FCC and FCW ones, which practically overlap. For all magnetic fields the magnetisation increases monotonically with decreasing temperature with the exception of  40~mT, where for both orientations a broad maximum occurs around 30~K, that corresponds to $B_{c1}$. Close to $T_c$ all data follow a generic curve that is reached at $B=100$~mT for $B\parallel \langle100\rangle$ and beyond $B=120$~mT, e.g. at 150~mT for $B\parallel \langle110\rangle$. The larger magnetic field value for $B\parallel\langle110\rangle$ should be attributed to the magnetocrystalline anisotropy. Deviations from these generic curves occur at temperatures and fields that correspond to the $B_{c2}$ values determined from the inflection points of $\Delta M/\Delta B$ as derived from Fig.~\ref{Fig:MvsB.pdf}(b). For instance, for $B\parallel \langle100\rangle$ at 80~mT $M$~deviates from the generic curve below $\sim$45~K, and this temperature shifts to higher values for lower magnetic fields. 

These generic curves are the same for both sample orientations and can be accounted for by a modified power law \cite{Zivkovic:2012kr} illustrated by the continuous lines in Fig.~\ref{Fig:MvsT.pdf}:
\begin{equation}
\label{eq:msat}
 M_{sat}(T) = M_{sat}(0) \; \left[1-\left(T/T_c^0\right)^n \right]^\beta
 \end{equation}
\noindent with $M_{sat}(0)\approx0.54$~$\mu_B$/Cu, $n\approx2$, $\beta\approx0.38$ and $T_c^0\approx59.7\;\text{K} \sim T_c+1.5\; \text{K}$. The exact values of the parameters are given in Table \ref{Tbl:powerlaw} and are in agreement with a previous study \cite{Zivkovic:2012kr}. Close to $T_c$, Eq.~\ref{eq:msat} reduces to the simple power law (for $n=1$) predicted theoretically \cite{Janson:2014uo}, that mimics a 3D-Heisenberg behaviour. 
%############################################################################
%############################################################################

%############################################################################
%    figures
%   chivsB_40_5K.pdf#
%############################
\begin{figure*}
\includegraphics[width= 0.7\textwidth]{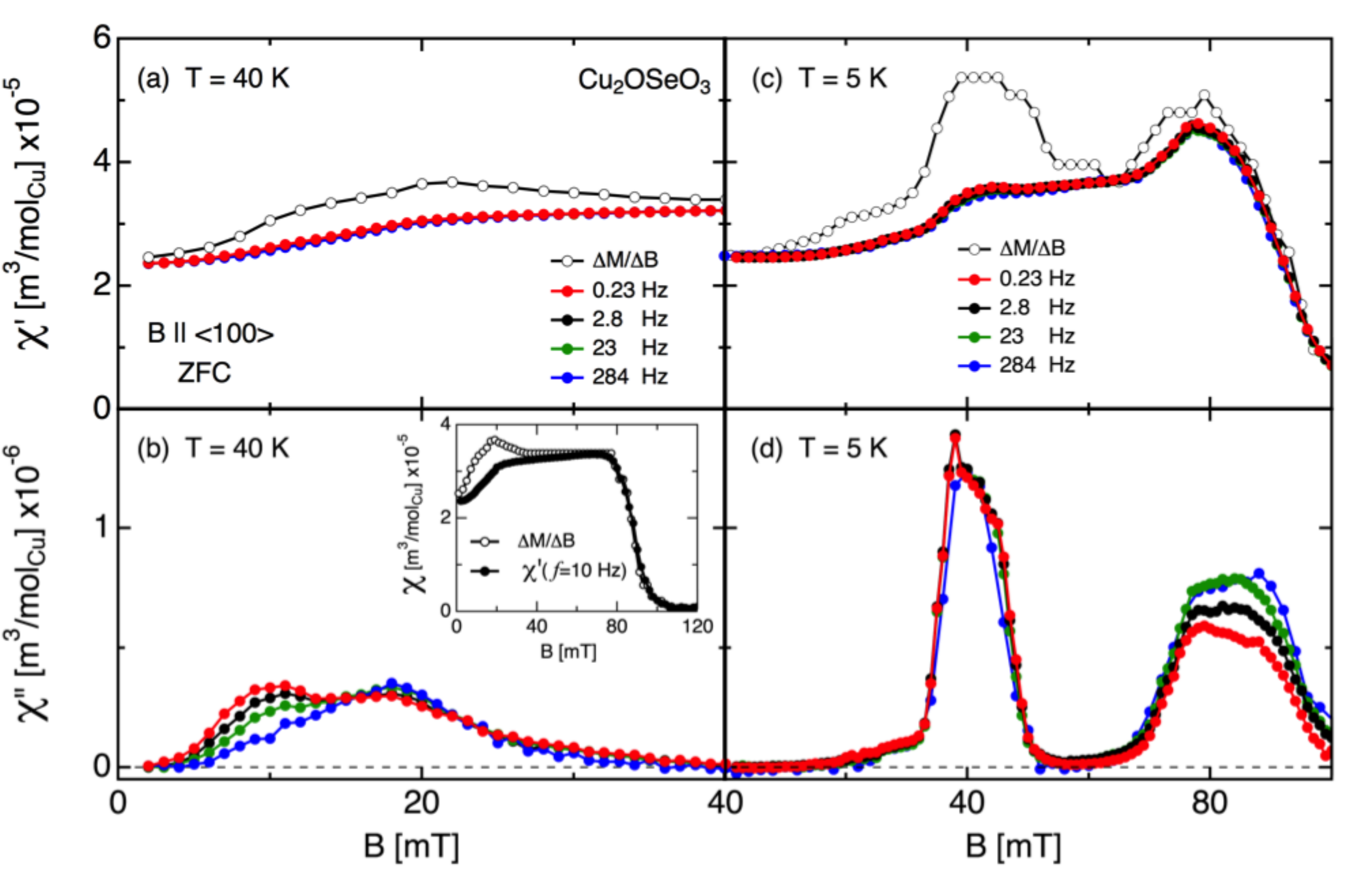}
\caption{ZFC $\chi'$ and $\chi''$ as a function of magnetic field at various frequencies for 40~K in panels (a) and (b), and 5~K in panels (c) and (d) with $B\parallel \langle100\rangle$. The dc susceptibility $\Delta M/\Delta B$ from Fig.~\ref{Fig:MvsB.pdf}(b) is included as well and extends beyond $B_{c2}$ up to 120~mT for 40~K in the inset of (b).} 
\label{fig:405K.pdf}
\end{figure*}
%############################################################################

%############################################################################
%    figures
%   chivsf_5K.pdf#
%############################
\begin{figure}
\includegraphics[width= 0.45\textwidth]{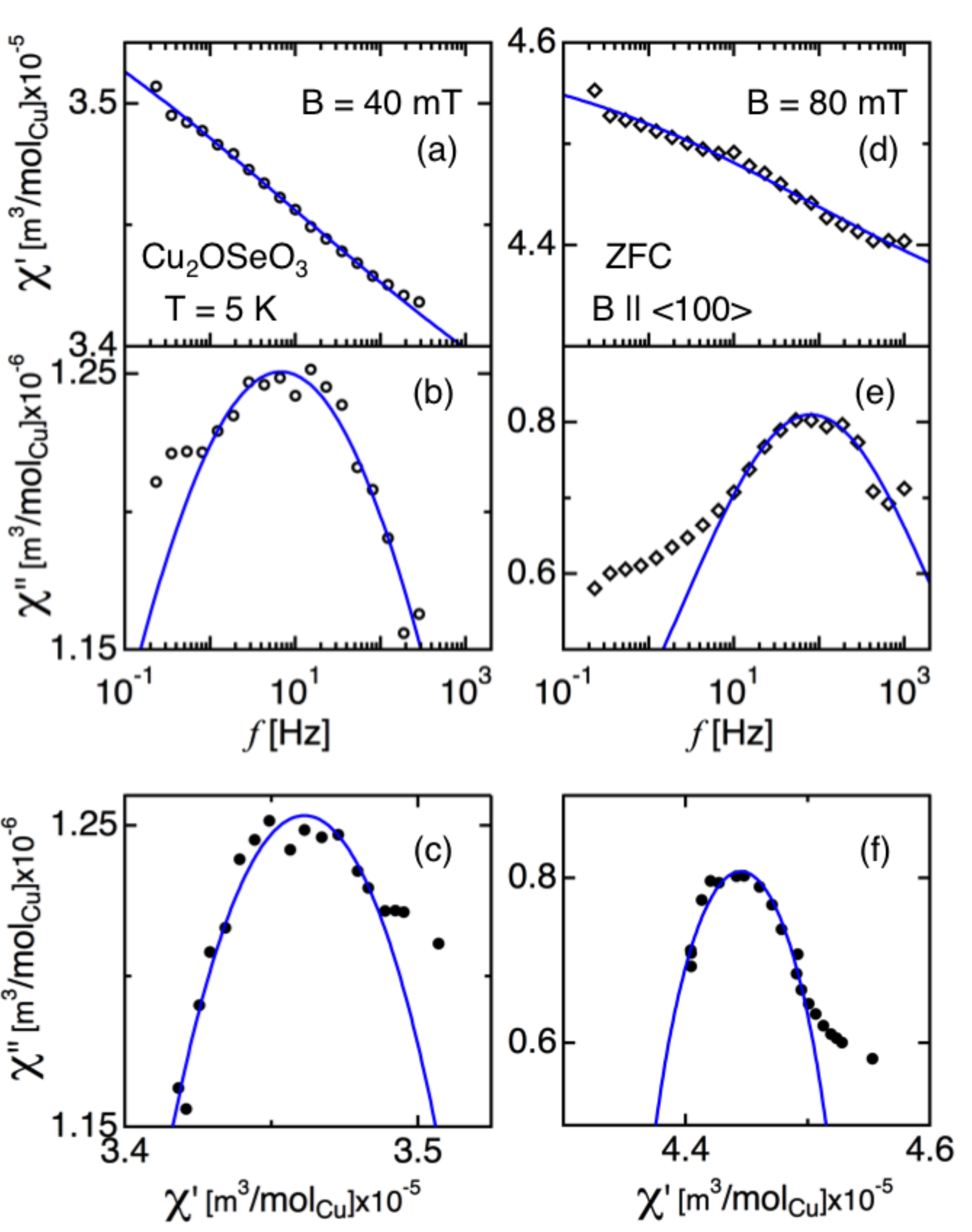}
\caption{Frequency dependence of $\chi'$ and $\chi''$ at (a and b) 40~mT and (d and e) 80~mT for $B\parallel \langle100\rangle$. Panels (c) and (f) show $\chi''$ plotted versus $\chi'$ for 40 and 80~mT respectively. The solid lines represent  fits of the Cole-Cole formalism to the data.} 
\label{fig:chivsf_5K.pdf}
\end{figure}
%############################################################################
%############################################################################
%    figures
%   PhaseDiagram.pdf#
%############################
\begin{figure*}
\includegraphics[width= 0.9\textwidth]{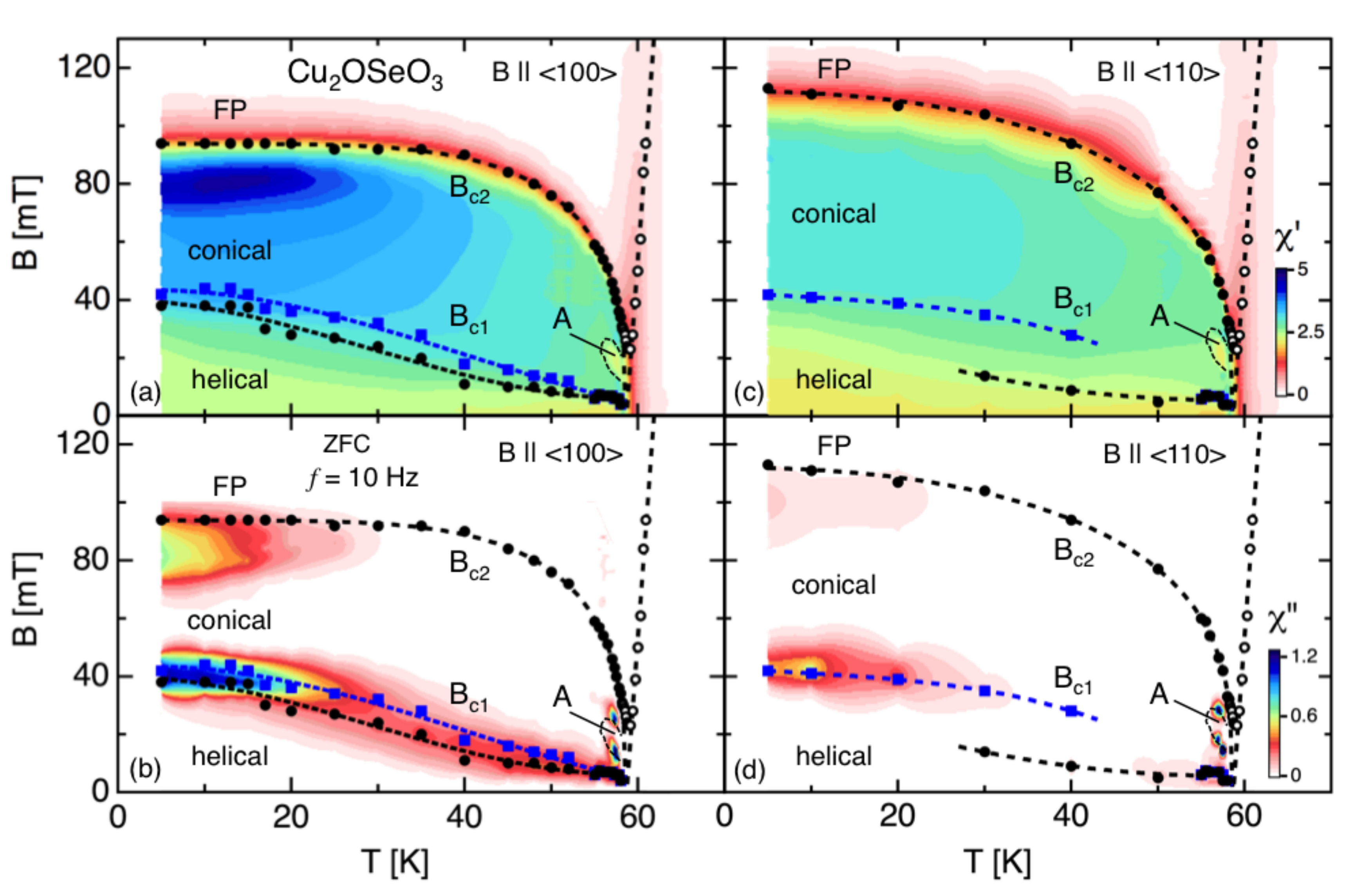}
\caption{Contour plots of $\chi'$ and $\chi''$ displayed versus temperature and magnetic field at $f$ = 10 Hz from ZFC measurements for $B\parallel\langle100\rangle$ in panel (a) and (b), and $B\parallel\langle110\rangle$ in panel (c) and (d). The phase boundaries $B_{c1}$, $B_{A1}$, $B_{A2}$ and $B_{c2}$ are shown as well. Below $\sim$50~K two $B_{c1}$ lines are illustrated by the circles and squares at lower fields. In the contour plots, we identify the following phases: helical, conical, A for the A-phase and FP for the field polarised one. The units for $\chi'$ and $\chi''$ are 10$^{-5}$ and 10$^{-6}$m$^3$/mol$_\text{Cu}$ respectively.} 
\label{fig:PhaseDiagram.pdf}
\end{figure*}
%############################################################################

\section{ac  susceptibility at 10~$\text{Hz}$}
\label{sec:ac}

Figure~\ref{fig:chivsB.pdf} illustrates the magnetic field dependence of ZFC $\chi'$ and $\chi''$ at an intermediate frequency of $f=10$~Hz for selected temperatures between 5 and 50~K. For both $B\parallel\langle100\rangle$ and $\langle110\rangle$  $\chi'$ increases slowly at low fields, similarly to  dc susceptibility $\Delta M/\Delta B$. On the other hand,  $\chi'$  in Figs.~\ref{fig:chivsB.pdf}(a) and \ref{fig:chivsB.pdf}(c) does not show the peak at $B_{c1}$ found in the $\Delta M/\Delta B$ curves, but a crossover to a plateau-like behaviour, characteristic of the conical phase. At 50~K, the plateau is very much horizontal for both field configurations, but at lower temperatures a slope sets in and below 30~K a peak develops at $\sim$ 80~mT, which is very weak for $B\parallel\langle110\rangle$ but well pronounced for $B\parallel\langle100\rangle$.

The corresponding $\chi''$ is shown in Figs.~\ref{fig:chivsB.pdf}(b) and \ref{fig:chivsB.pdf}(d)  for both orientations. Around $B_{c1}$, for $B\parallel\langle100\rangle$, $\chi''$ seems to consist of two adjacent peaks, which at 50~K are centred at 8 and 13~mT respectively.  With decreasing temperature the amplitude of these peaks increases considerably and their positions move to higher magnetic fields. At 5~K they occur at 38 and 42~mT respectively. These two peaks are also observed for $B\parallel \langle110\rangle$ but with relatively weaker intensity and occur simultaneously only between 40 and 30~K. Above $B_{c1}$ for both orientations, $\chi''$ is zero in the conical phase, thus no dissipation is observed. 

However close to $B_{c2}$, a broad maximum of $\chi''$ appears below 30 K, which is hardly visible for $B\parallel \langle110\rangle$ but prominent for $B\parallel \langle100\rangle$, a configuration where the peak of $\chi'$ develops as well. Similarly to the peak of $\chi'$ for $B\parallel \langle100\rangle$, the maximum of $\chi''$ increases with decreasing temperature. Such a peak in $\chi''$ has not been observed in  B20  magnets so far and seems to be a particularity of \cuse. 

The effect of thermal history is illustrated by Fig.~\ref{fig:chivsT.pdf}, which depicts the temperature dependence of $\chi'$ and $\chi''$ for FCC and FCW and $B\parallel \langle100\rangle$. A  pronounced hysteresis is observed for $\chi''$ at $B=40$~mT  below 40~K, which is hardly visible in $\chi'$  probably due to its weak temperature dependence in this temperature range. On the other hand, strong hysteresis appears in both  $\chi'$ and $\chi''$ at 80~mT below  30~K. At higher magnetic fields, e.g. 100~mT, both $\chi'$ and $\chi''$ are much weaker and they hardly show any hysteresis. 

%############################################################################
%############################################################################
\section{Frequency dependence of ZFC ac magnetic susceptibility}

The relaxation phenomena associated with $B_{c1}$ and $B_{c2}$ presented in the previous section are most visible for $B\parallel \langle100\rangle$ and in the following we will focus on the results obtained in this orientation. Figure~\ref{fig:405K.pdf}  displays the ZFC $\chi'$ and $\chi''$ for various frequencies at 40 and 5~K. The figure highlights the differences  between $\Delta M/\Delta B$ and $\chi'$ as $\Delta M/\Delta B$ develops a peak around $B_{c1}$ while $\chi'$ shows a smooth crossover to a plateau. Above $B_{c1}$, the ac and dc susceptibilities converge for both temperatures and overlap close to $B_{c2}$. 

The frequency dependence  is more pronounced in $\chi''$ in contrast to that of $\chi'$. At 40~K, $\chi''$ shows two peaks which are located at 10 and 18~mT, thus close to $B_{c1}$. Clear frequency dependence is observed in the amplitude of these peaks and is different for the two peak positions. The peak at 10~mT decreases with increasing frequency, while the one at 18~mT slightly increases. At 284~Hz only the peak at 18~mT is visible. On the other hand, at 5~K these two $\chi''$ peaks seem to merge as shown in Fig.~\ref{fig:405K.pdf}(d) and vary very weakly with frequency.  In contrast, the broad peak in $\chi''$, that appears close to $B_{c2}$, shows a noticeable frequency dependence.

A quantitative analysis of the frequency dependence of $\chi'$ and $\chi''$ is provided by the modified Cole-Cole formalism \cite{cole_cole, Hueser86}:
%############################################################################
\begin{equation}
\chi(\omega)= \chi(\infty)+\frac{\chi(0)-\chi(\infty)}{1+(i\omega \tau_0)^{1-\alpha}}
\label{eq:colecole_start}
\end{equation}
%############################################################################
\noindent with $\omega=2\pi f$ the angular frequency, $\chi(0)$ and $\chi(\infty)$ the adiabatic and isothermal susceptibilities respectively and $\tau_0=1/(2\pi f_0)$ the characteristic relaxation time with $f_0$ the characteristic frequency. The parameter $\alpha$ measures the distribution of characteristic relaxation times: $\alpha=0$ corresponds to a single relaxation time, and $\alpha>0$ to a distribution of relaxation times, which becomes broader as $\alpha$ approaches 1. Equation~\ref{eq:colecole_start} leads to the in-phase and out-of-phase components of the susceptibility, and a relation between these two quantities  discussed in our previous report  \cite{Qian2016}. 

Figure~\ref{fig:chivsf_5K.pdf} illustrates the frequency dependence of ZFC $\chi'$ and $\chi''$ at 5~K for 40 and 80~mT. With increasing frequency, $\chi'$ decreases very weakly, only by $\sim$3\%, as seen in the panels (a) and (d).  A more significant variation is found for $\chi''$, which shows broad maxima at the characteristic frequencies $f_0\sim10$~Hz and 100~Hz for 40 and 80~mT, thus around $B_{c1}$ and $B_{c2}$ as seen in the panels (b) and (e), respectively. 

%The interrelation between $\chi'$ and $\chi''$ is given in Figs.~\ref{fig:chivsf_5K.pdf}(c) and (f).  

The fit of eq.~\ref{eq:colecole_start} to the data is illustrated by the solid lines in Fig.~\ref{fig:chivsf_5K.pdf}. It leads to $\alpha\sim0.85$ and $\sim0.65$ for $B$=40 and 80~mT, respectively, thus revealing the existence of broad distributions of relaxation times. In addition,  deviations from the simple Cole-Cole behaviour are found at low frequencies, as illustrated  in panels (b), (c), (e) and (f). These may originate from the co-existence of several relaxation processes similarly to what has been reported at $B_{c1}$ close to $T_c$ \cite{Qian2016} and over a larger temperature and magnetic field range in Fe$_{1-x}$Co$_x$Si \cite{Bannenberg:2016wf}. 
%############################################################################
 %############################################################################%
%## power Law
%############################
\begin{table*}
\centering
\caption{Parameters of the power law fits to the temperature dependence of $M_{sat}$ and $B_{c2}$ below $T_c=58.2$~K for $B\parallel\langle100\rangle$ and $B\parallel\langle110\rangle$. For the sake of comparison the corresponding parameters from the  literature are also provided.}
	\begin{tabular}{ p{2cm} | p{2.5cm} p{2.5cm} p{2.5cm} | p{2.3cm} p{2.3cm} p{2cm} }
%	\begin{tabular}{|| c | ccc | ccc || }
	\hline
	\hline
	 \multirow{2}{*}{} &\multicolumn{3}{c |} { $M_{sat}(T)$} &  \multicolumn{3}{c } {$B_{c2}$ } \\
	 & ~$B\parallel\langle100\rangle$ & $B\parallel\langle110\rangle$ & Ref[\citenum{Zivkovic:2012kr}]&~$B\parallel\langle100\rangle$ & $B\parallel\langle110\rangle$ & Ref[\citenum{Levatic:2016ed}] \\
	\hline
	~$M_{sat}(0)$ & ~0.54$\pm$0.01 & 0.53$\pm$0.01 & 0.559$\pm$0.007  &~ -- & -- & -- \\
	~$B_{c2}(0)$ &  ~-- & -- & -- & ~94$\pm$2 & 112$\pm$1 & 85 \\
	~$T_c^0-T_c$ &~1.5$\pm$0.1 & 1.8$\pm$0.3 & 1.8$\pm$0.1 & ~0.3$\pm$0.1 & 0.4$\pm$0.1 & 0\\
	~$\beta$ & ~0.368$\pm$0.005 & 0.391$\pm$0.002 & 0.393$\pm$0.004 &~0.34$\pm$0.02 & 0.30$\pm$0.01 & 0.25\\
	~$n$  &~1.99$\pm$0.02 & 2.01$\pm$0.01 & 1.95 &~5.1$\pm$0.2 & 2.2$\pm$0.2 & 1\\
	\hline
	\hline
	\end{tabular}
\label{Tbl:powerlaw}
\end{table*}
%############################################################################
%############################################################################
\section{Discussion}

The results presented in the previous sections reveal new features and qualitatively different phase diagrams for $B\parallel\langle100\rangle$ and $\langle110\rangle$. The magnetic phase diagrams are illustrated as contour plots in Fig.~\ref{fig:PhaseDiagram.pdf}, which summarise the  magnetic field and temperature dependence of the ZFC $\chi'$ and $\chi''$ at $f=10$~Hz for $B\parallel\langle100\rangle$ and $\langle110\rangle$ and include additional data close to $T_c$ from our previous study \cite{Qian2016}. The phase boundaries  shown are determined following the same criteria as in our previous study \cite{Qian2016}: $B_{c1}$, $B_{A1}$ and $B_{A2}$ from the peaks of $\chi''$, and $B_{c2}$ from the inflection points of $\chi'$ vs $B$. 

We  start  the discussion with the behaviour of $\chi''$ close to $B_{c1}$, which shows  two adjacent peaks  below $\sim$50~K, for $B\parallel\langle100\rangle$,  as seen in Figs.~\ref{fig:chivsB.pdf}(b),~\ref{fig:405K.pdf}(b) and~\ref{fig:405K.pdf}(d).  These peaks result in  two $B_{c1}$ lines that merge into a single broad peak just below $T_c$,  the frequency dependence of which bears the signature of more than one relaxation process \cite{Qian2016}.  On the other hand,  the peaks of $\chi''$ for  $B\parallel\langle110\rangle$ lead to  two discontinuous $B_{c1}$ lines, which coexist  only at a limited temperature range, between $\sim$40 and $\sim$30~K, as seen in Fig.~\ref{fig:chivsB.pdf}(d). This complex behaviour for both field orientations seems to be unique to \cuse~and might be related to the two different Cu$^{2+}$ sites, which might lead to slightly different magnetocrystalline anisotropies. 

At high magnetic fields, a single  $B_{c2}$ line is found for each of the sample orientations. Close to $T_c$, the temperature dependence of $B_{c2}$ is similar for both orientations but differences appear below $\sim50$~K: $B_{c2}$ levels off with decreasing temperature  for $B\parallel\langle100\rangle$, while it keeps increasing for  $B\parallel\langle110\rangle$. The temperature dependences of the two $B_{c2}$ lines can be accounted for by Eq.~\ref{eq:msat} and lead to the parameters  given in Table~\ref{Tbl:powerlaw} with $\beta\approx0.32$ and $T_c^0\approx58.6$~K$\sim T_c+0.4$~K for both orientations and  different $n$: $n=5.1\pm0.2$ for $B\parallel\langle100\rangle$ and $n=2.2\pm0.2$ for $B\parallel\langle110\rangle$. These values deviate from those reported in the literature \cite{Levatic:2016ed}  and from those deduced from the temperature dependence of $M_{sat}$. On the other hand, close to $T_c$, both $B_{c2}$ lines reduce to the same simple power law, for $n= 1$, that has been previously reported for the same samples \cite{Qian2016}. 

%but with parameters very different from those deduced from the $M_{sat}$ versus $T$ curves
%The larger $n$ for $B\parallel\langle100\rangle$ may be attributed to the appearance of the peak in $\chi'$ close to $B_{c2}$, which lowers the magnetic field values of the inflection points used for defining $B_{c2}$.

Besides the different phase boundaries, the two sample orientations also lead  to substantially different behaviour for both $\chi'$ and $\chi''$ close to $B_{c2}$ and below $\sim$30~K. The maxima that appear both in $\chi'$ and $\chi''$ in this field and temperature range are marked for $B\parallel\langle100\rangle$, but are much weaker for $B\parallel\langle110\rangle$, as  shown in Fig.~\ref{fig:chivsB.pdf}. Similar differences are also found in the dc magnetisation as well as in  previous reports \cite{Zivkovic:2012kr, Belesi:2012fi, Adams:2012du}. Theory  predicts a similar behaviour  but with much more dramatic effects \cite{Janson:2014uo}. Indeed, while a continuous second order transition is expected when a magnetic field is applied outside the easy magnetisation axis, a discontinuous first order transition is predicted  for  $B \parallel \langle100\rangle$, i.e. when the magnetic field is applied along the easy magnetisation axis. This first order transition should appear as a jump of the magnetisation at $T=0$~K, leading to a divergence of the derivative, thus of the susceptibility. According to theory the jump of the magnetisation at $B_{c2}$ should be visible  up to  $\sim$50~K. In  our measurements the maxima in $\chi'$ and $\Delta M/\Delta B$ appear only below 20~K. Thus, there is a  limited  agreement with theory at $B_{c2}$, which however, does not expand to $B_{c1}$, where  theory predicts a single transition line and  does not account for the two lines observed experimentally.

At this moment, we are not aware of any theory that could explain  the significant dissipation phenomena that appear  in $\chi''$ and occur not only at $B_{c1}$ but also at  $B_{c2}$ at low temperatures. In particular the strong $\chi''$ at $B_{c2}$ for $B \parallel \langle100\rangle$ has not been reported before neither for \cuse~nor for other B20 chiral magnets and seems to be unique to \cuse. A discussion on the origin of these effects can only be speculative based on our main observations that are  the dependence on the sample orientation with respect to the magnetic field,  the broad distributions of relaxation times seen in the values of $\alpha$, the relatively long characteristic relaxation times $\tau_0$ and the coexistence of several relaxation processes. Among these, the dependence on the sample orientation at $B_{c2}$ might be related to the effects seen in the magnetisation curves and the theoretical predictions discussed in the previous paragraph. 

The characteristic  long relaxation times deduced from the Cole-Cole analysis close to  $B_{c1}$ and  $B_{c2}$ indicate rearrangements of large correlated volumes with extremely low associated energies of the order of $\hbar\omega$. In addition, it seems reasonable to assume that the distribution of relaxation times also reflects a distribution of volumes.  It is not clear what  the size of these volumes is and what delimits them, possibly defects, stress or frustration.  These very slow relaxation phenomena might prevent the system from reaching thermodynamic equilibrium at low temperatures and thus explain the hysteresis between FCC and FCW data seen in Fig.~\ref{fig:chivsT.pdf}.

%The characteristic  long relaxation times deduced from the Cole-Cole analysis indicate that the transitions at  $B_{c1}$ and  $B_{c2}$ involve rearrangements of large correlated volumes. However it seems reasonable to assume that the distribution of relaxation times and the coexistence of several relaxation processes reflect a distribution of volumes and possibly a superposition of several volume distributions. I
%############################################################################
%############################################################################
\section{Conclusion}

The dc magnetisation and ac susceptibility of \cuse~reveal qualitatively different phase diagrams below 50~K depending on whether the  magnetic field was applied  along  the easy  $\langle100\rangle$  or  the hard  $\langle110\rangle$  crystallographic directions of the sample. The differences are not only in the specific values of the critical fields  $B_{c1}$ and  $B_{c2}$ but also in the  behaviour of the magnetisation and  susceptibility.

The $B_{c1}$ transition from the helical to the conical phase splits in two well distinct lines below $\sim$50~K, which may be related to the two different Cu$^{2+}$ sites. These two  $B_{c1}$ lines  coexist only between 30 and 40~K for $B \parallel \langle110\rangle$, but  expand almost over the whole temperature range for  $B \parallel \langle100\rangle$.  
 
Close to $B_{c2}$, the magnetisation and the susceptibility are in qualitative agreement with theory, which however foresees  more dramatic effects with a discontinuous first order transition from the conical to the field polarised state when the magnetic field is applied along the easy axis and a continuous second order transition otherwise. In addition,  below 30~K, an important $\chi''$ signal appears close to $B_{c2}$ for $B \parallel \langle 100\rangle$, the origin of which is unclear.  Similar dissipation phenomena have not been observed in \cuse~or in other B20 magnets before.
 
At both $B_{c1}$ and  $B_{c2}$  the frequency dependence of $\chi''$ reflects the existence of very broad distributions of relaxation times and the superposition of several relaxation processes, which lead to thermal hysteresis  and eventually prevent the system from reaching  thermal equilibrium at low temperatures.  
%This study provides detailed dc magnetisation and ac susceptibility measurements of \cuse~at low temperatures. A key point of this study is a clear observation of $\chi'$ and $\chi''$ signal close to $B_{c2}$ for the first time. This signal occurs only for $B\parallel\langle100\rangle$, and has a characteristic frequency of $\sim100$~Hz. A clear thermal history effect is seen from both $\chi'$ and $\chi''$ below 30~K. In addition, the transition between the helical and conical phases shows two $B_{c1}$ lines below 55~K with different frequency dependence, which may be related to the two Cu$^{2+}$ sites in \cuse.

\begin{acknowledgements}
The authors would like to thank G.R. Blake for checking the sample orientations and L.J. Bannenberg for fruitful discussions. F. Qian acknowledges  financial support from the China Scholarship Council. 
\end{acknowledgements}
\newpage

\bibliography{Ref_CuSe_suscep_lowT}

\end{document}